# Threshold Voltage Jitter due to Random Telegraph Noise

Gilson Wirth

Department of Electrical Engineering, Univ. Federal do Rio Grande do Sul – UFRGS

Porto Alegre, RS, Brazil

e-mail: gilson.wirth@ufrgs.br

*Abstract*— **With the downscaling of MOSFETs to nanometer dimensions, transistor electrical parameter variability is produced by factors other than variations of physical dimensions and doping profiles, which are there since device fabrication and remain static over time. Besides these time-zero variability factors, factors that lead to performance variability from one instant in time to the other start playing a significant role. Random Telegraph Noise (RTN) is among these relevant time-dependent variability sources. In this work we extend the knowledge of the time-dependent random variability induced by RTN, by providing a statistical model for transistor threshold voltage jitter $\sigma_{\Delta V_T}$ produced by RTN. The area scaling of $\sigma_{\Delta V_T}$ is detailed and discussed, supporting designers in transistor sizing towards a more reliable design. Not only the jitter expected in a transistor is modeled, but also its variability among transistors that by design should be equal. Besides analytical modeling, Monte Carlo simulations are run. The simulations account for the charge carrier capture and emission events related to RTN, allowing the proper evaluation of the RTN related jitter. The Monte Carlo simulations validate the analytical model and illustrate the area scaling of jitter and its variability.**

*Index Terms*—**metal–oxide–semiconductor field-effect transistor (MOSFET) scaling, random telegraph noise (RTN), timing jitter, time-dependent variability.**

## 1. Introduction

MOSFETS are employed in digital, analog and mixed-signal integrated circuit (IC) designs. Yield and reliability of ICs using MOSFTES depend on the variability of the transistor parameters and noise. Both stochastic variability and noise scale inversely with area. On the other hand, cost increases with area, and increasing area may also increase capacitive load, what decreases performance and increases power consumption. For the designer to be able to find the adequate balance between cost, reliability and performance, adequate models are demanded.

There are sources of variability that are time independent, present in the fresh (new) device due to imperfection in fabrication process and the discrete nature of the matter. This includes line edge roughness, random dopant fluctuations, and metal (or poly) gate granularity. But there is also time dependent - time varying - sources of variability. Random Telegraph Noise (RTN) is among these time varying sources of variability [1, 2].

There are well stablished models for the time-zero variability. To a first order, time-zero parameter variability (parameter statistical standard deviation) is considered to be inversely proportional to the square root of the area [1]. Furthermore, in circuits that contain a large number of small area transistors, the impact of RTN-induced fluctuation is considered to increase when it is compared with the static variation caused by manufacturing process [2].

In this work we discuss and detail the area dependence of the time dependent variability induced by RTN.

In digital circuits, the RTN chronological statistics, especially trap occupancy switching, has direct impacts on circuit performance and reliability, as degradations like jitter of signals happen when a trap switches state.

In this work, an analytical model to evaluate the RTN induced threshold voltage jitter RTN and its impact in digital circuits is presented. To the best of our knowledge, this is the first statistical model derivation of the RTN induced threshold voltage jitter and its variability. Properly modeling variability is of paramount relevance, since not only the amplitude of RTN increases with area downscaling, but also the variability increases. Variability increases even faster than average amplitude, as shown below.

## 2. ANALYTICAL MODEL

### A. Model Derivation

The alternate capture and emission of carriers at individual defect sites (traps) generates discrete fluctuations in the device current. These fluctuations, also called Random Telegraph Noise (RTN), are the main source of Low-Frequency Noise in deep-submicron MOSFETs.

Figure 1 shows an RTN time trace measured on a small area nMOSFET, originating from a single trap. The current is seen to alternate between a higher current state and a lower current state. The difference (fluctuation) in current between states is $\delta I_D$. In large area transistors a large number of traps is usually found, but with a reduced individual impact. Usually current fluctuations are measured. The current fluctuation $\delta I_D$ may be translated into a threshold voltage fluctuations $\delta V_T$ [1, 3, 7, 11-12].

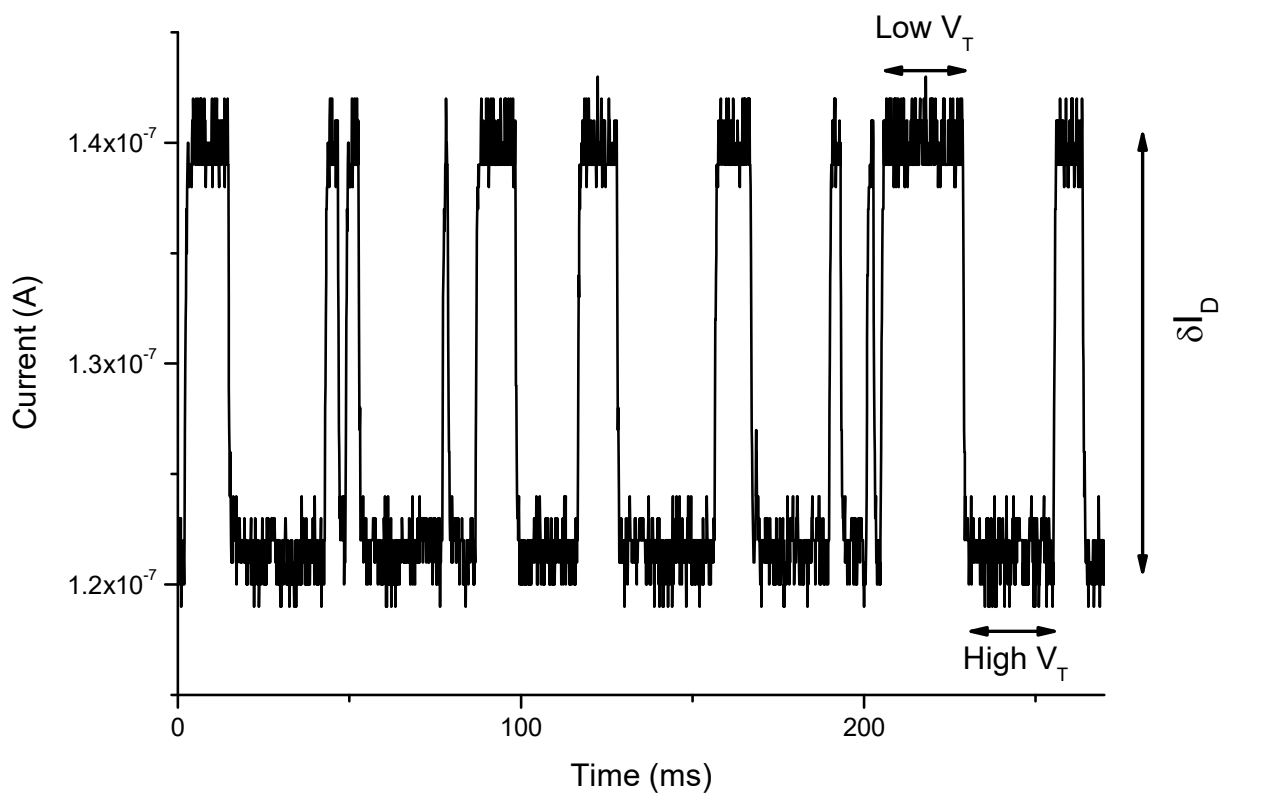

Fig. 1. RTN measurement in time domain. Charge carrier capture and emission by traps lead to discrete fluctuations in drain current over time. The high current state, which corresponds to the state where the trap is electrically neutral (empty), refers to the lower $V_T$. The low current state, which corresponds to the state where the trap is electrically charged, refers to the higher $V_T$. $\delta I_D$ is the amplitude of the current fluctuation.

This capture and emission of electrons by a charge trap may be modeled as a two-state fluctuation of the threshold voltage $V_T$. If the trap is empty, we consider that the device is at a lower $V_T$ (hence higher current). If a charge carrier is trapped, we consider the device is at a higher $V_T$ (hence lower current). The difference in $V_T$ between the trapped and empty state of a single trap is then $\delta V_T$. Considering the *ith* trap, $\delta V_{Ti}$ is the $V_T$ fluctuation due to the *ith* trap. Please note that for traps that lead to a higher $V_T$ if occupied (and hence a lower $V_T$ if empty) the notation for the states may be exchanged, without any loss of generality and model validity.

The value of the $V_T$ fluctuation due to all traps at time *t*, here called $\Delta V_T(t)$, is then evaluated as the sum of the contribution of all *n* traps found in a device:

$$\Delta V_T(t) = \sum_{i=1}^{n} \delta V_{Ti}\, S_i(t) \qquad (1)$$

For simplicity and consistency with noise, we make $\langle \Delta V_T(t) \rangle = 0$ (noise is assumed to have an average value of zero). This is also convenient to lead to an elegant resolution for the jitter and its variability. Zero average value is achieved by making the average value of the $V_T$ fluctuation due to each trap equal to zero, by writing $S_i(t)$ as being

$$S_i(t) = \begin{cases} -1.P_i(1) & \text{if the trap is empty} \\ +1.P_i(0) & \text{if the trap is occupied} \end{cases} \qquad (2)$$

Where $P_i(1)$ is the probability of the *ith* trap being occupied and $P_i(0)$ is the probability of the *ith* trap being empty:

$$P_i(0) = \frac{\tau_{Ci}}{\tau_{Ci}+\tau_{Ei}} \text{ and } P_i(1) = \frac{\tau_{Ei}}{\tau_{Ci}+\tau_{Ei}} \qquad (3)$$

where $\tau_C$ and $\tau_E$ capture and emission time constants, respectively.

Please note that this notation does not change the waveform of the RTN. It just removes the average value (DC component) of the waveform. The amplitude of the $\Delta V_T(t)$ induced by the the *ith* trap keeps being $\delta V_{Ti}$, since $P_i(0)+P_i(1)=1$. See Fig. 2(b), where this is exemplified. The average value (DC component) of the waveform does not contribute neither to jitter nor to noise. Hence, it is appropriate to remove it, besides facilitating the statistical analysis.

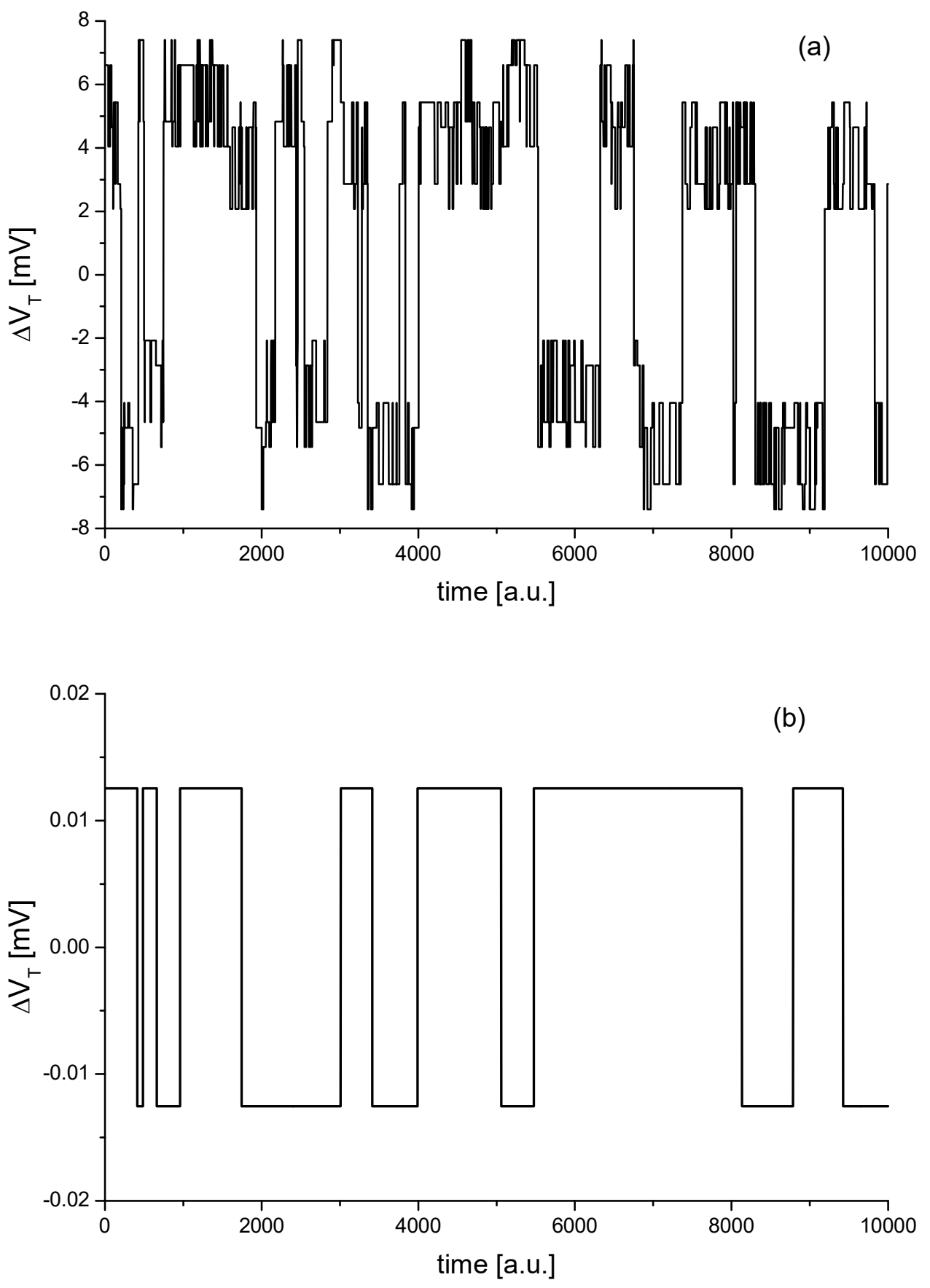

Fig. 2. Monte Carlo simulation of threshold voltage variation over time *ΔV<sub></sub>*$\Delta V_T(t)$ due to Random Telegraph Noise (RTN). Trap occupancy switching leads to discrete fluctuations in $V_T$. The first $10^4$ points of the Monte Carlo runs for two devices of size 0.03µm x 0.12µm are show. Due to random number of traps and related parameters, devices that by design should be identical show different jitter.

Then the variance of the threshold voltage fluctuation is evaluated starting from

$$\sigma_{\Delta V_T}^2 = < \Delta V_T(t)^2 > - < \Delta V_T(t) >^2 \qquad (4)$$

with $\langle \Delta V_T(t) \rangle = 0$.

For evaluating $\langle \Delta V_T(t)^2 \rangle$, we note that for a single trap

$$< (\delta V_{Ti} S_i(t))^2 > = \delta V_{Ti}{}^2 . \beta_i/(1+\beta_i)^2 = < A_i{}^2 > \qquad (5)$$

where $\beta_i = \tau_{Ci}/\tau_{Ei}$.

$\langle \Delta V_T(t)^2 \rangle$ is then evaluated as being

$$< \Delta V_T(t)^2 > = < \sum_{i=1}^{n} (\delta V_{Ti}\, S_i(t))^2 > = < n > < A_i{}^2 > \qquad (6)$$

which leads to

$$\sigma_{\Delta V_T}^2 = < n > < A_i{}^2 > \equiv V_{Tjitter}^2 \qquad (7)$$

In the case of $\tau_{Ci}=\tau_{Ei}$ (50% occupation probability) $A_i$ will be equal to $\delta V_{Ti}/2$. As in the case of noise, this ($\tau_{Ci}=\tau_{Ei}$) is the situation in which a trap generates the largest jitter. In the case of $\tau_{Ci}\neq\tau_{Ei}$, $A_i$ will be smaller than $\delta V_{Ti}/2$.

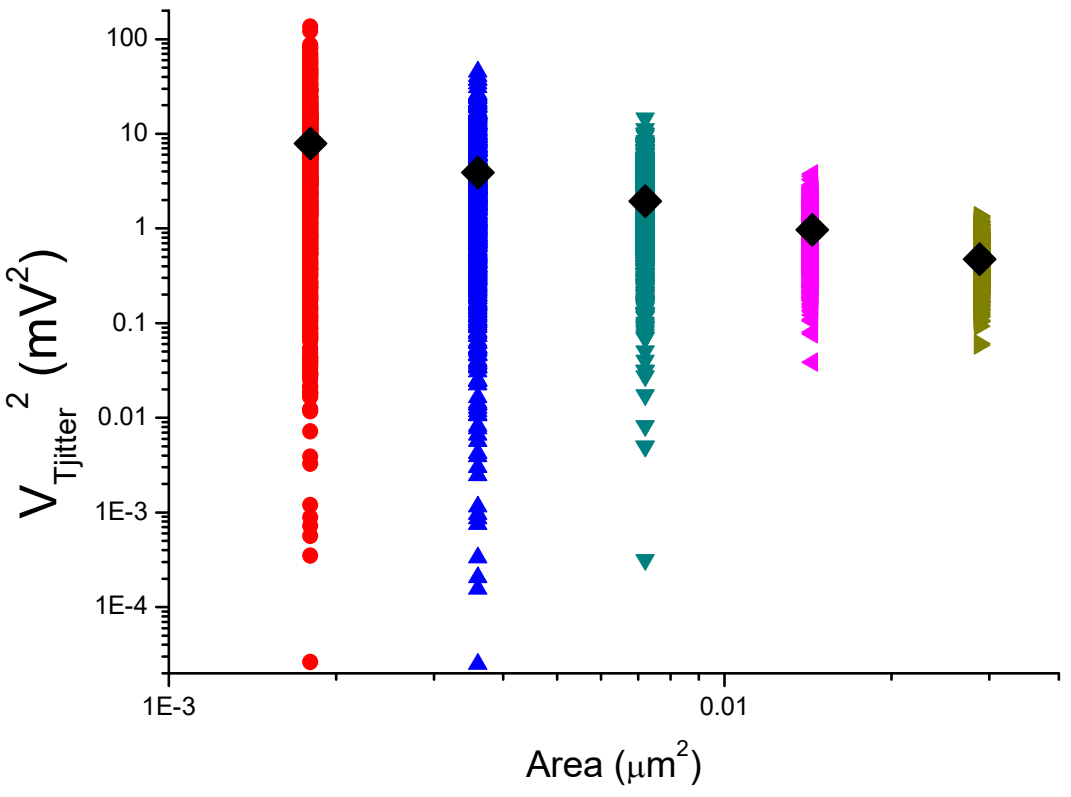

Fig. 3. $\sigma^2_{\Delta V_T} \equiv V^2_{Tjitter}$ as a function of device area. The simulated device sizes (WxL) are 0.06μm x 0.03 μm, 0.12 μm x 0.03 μm, 0.24 μm x 0.03 μm, 0.48 μm x 0.03 μm and 0.96 μm x 0.03 μm. Monte Carlo simulations are run with 1000 runs for each device size. Each point in the graph is the value for a Monte Carlo simulation run, i.e., the value of $\sigma^2_{\Delta V_T}$ of a single device. Black diamonds show the average jitter for each area, which is in good agreement to (7). Not only $V_T$ jitter average value increases with device downscaling, but also its spread (variability) increases.

Equations (6) and (7) above are for the average value of the jitter expected in a single transistor, i.e. the expected threshold voltage variation over time in a given transistor. It is also important to evaluate the variability of the expected (average) value of jitter among different transistors. Different transistors will show different jitter. Jitter variability may be evaluated using

$$\sigma^2_{V_{Tjitter}} =< {V^2_{Tjitter}}^2 > -< V^2_{Tjitter} >^2 \quad (8)$$

It is evaluated as being

$$\sigma^2_{V_{Tjitter}} =< n >< {A_i}^4 > \quad (9)$$

*B. Area Scaling and Relation to 1/f noise*

To detail the area scaling of $\sigma^2_{\Delta V_T}$ and $\sigma^2_{V_{Tjitter}}$ we must look at the area dependence of <n> and $A_i$. These are parameters well studied in the literature. The number of traps is known to be proportional to area, <n>~WL. The average amplitude to a trap is known to scale inversely with area, $A_i$~1/WL [1, 3, 7 and 8]. From (7) following relation between device area and average expected jitter value can then be written

$$V^2_{Tjitter} = \sigma^2_{\Delta V_T} \sim {}^{1}/_{WL} \quad (10)$$

For the variation of jitter among devices, from (9) following relation between device area and jitter variability among devices can be written

$$\sigma^2_{V_{Tjitter}} \sim {}^{1}/_{(WL)^3} \quad (11)$$

Please note that then $\sigma_{V_{Tjitter}} \sim {}^{1}/_{(WL)^{1.5}}$

This means that with area scaling not only the expected jitter value increases, but also the variability of expected jitter value increases. This increasing time-dependent random variability is a significant challenge for the device designer.

The average value of threshold voltage jitter is related to the average value of 1/f noise power. Equation (7) above has the same form and same area dependence as equation (34) in [3]. Both equations (7) above and (34) in [3] are proportional to $< {A_i}^2 >$ and proportional to the average number of traps in a given device size. The same area dependency is seen in (10) above and (35) in [3]. This is expected, since average value of jitter (and average value of phase noise) due to RTN is related to the average 1/f noise due to RTN. Also, similarly to 1/f noise (frequency domain) traps that have maximum contribution to jitter (time domain) are the ones with capture time similar to the emission time, i.e., $\beta \approx 1$.

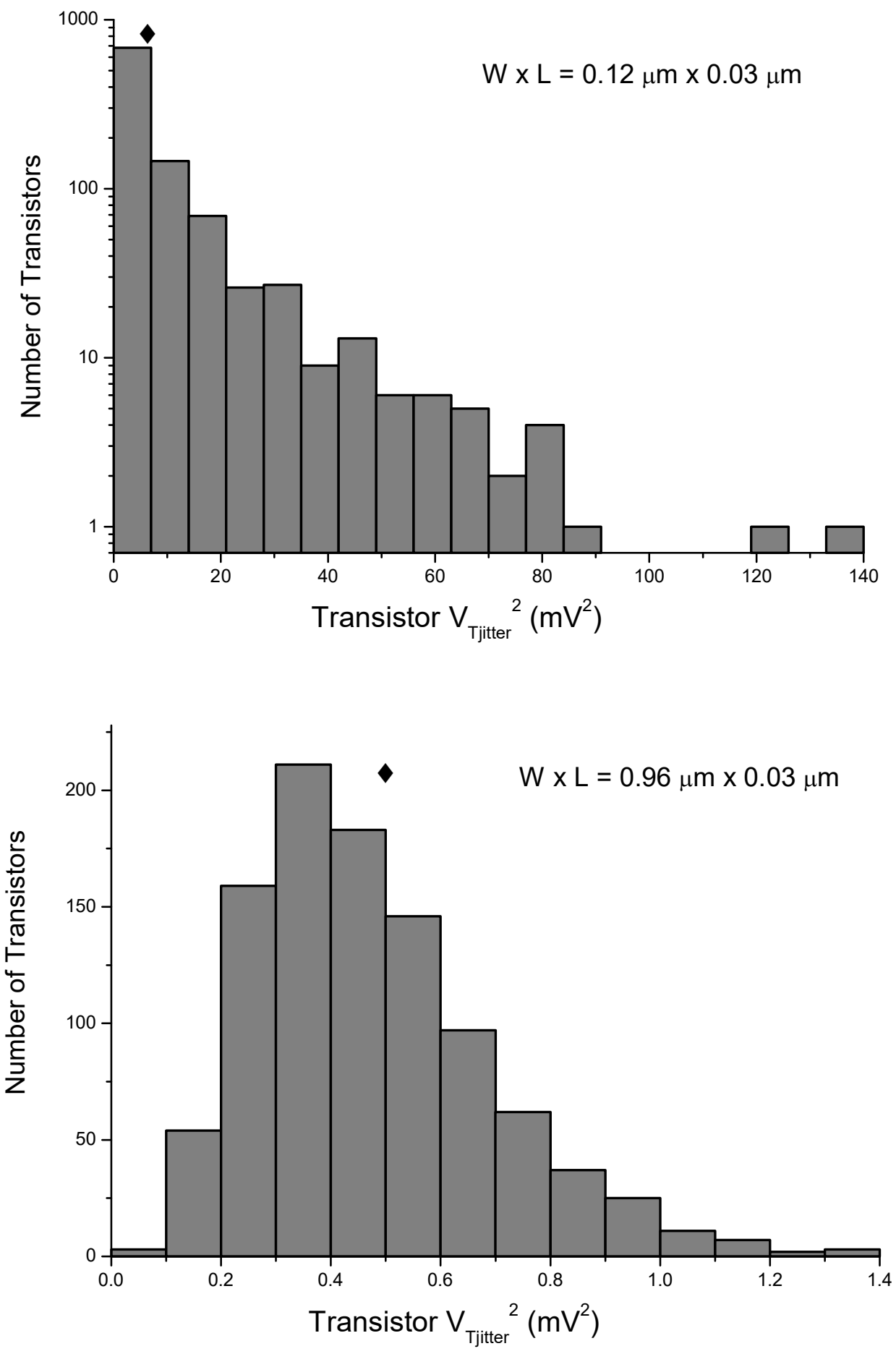

Fig. 4. Histogram of $V_{Tjitter}{}^2$ from Monte Carlo simulations for two different device sizes. For each size, 1000 devices were simulated. Black diamonds show the average $V_{Tjitter}{}^2$ for each device size. Please note that the y-axis of the first histogram is in log scale.

Similarly, the variability of threshold voltage jitter is related to the variability of 1/f noise power. Equation (9) above has the same form and same area dependence as equation (38) in [3]. Both equations (9) above and (38) in [3] are proportional to $< {A_i}^4 >$ and proportional to the average number of traps in a given device size. The same area dependency is seen in (11) above and (38) in [3]. This is expected, since jitter variability (and phase noise variability) due to RTN is related to the variability of 1/f noise due to RTN.

Hence, as seen in 1/f noise, not only the expected average value of jitter is expected to increase with device area downscaling, but also the variability of jitter performance between devices strongly increases with device downscaling. Each device has a random number of traps with random amplitudes and random time constants. The $V_T$ jitter of a device is given by its particular number of traps and related parameters. Number of traps and related parameters vary among devices. The smaller the device size, the larger the jitter variability among devices.

## 3. Monte Carlo Simulations

We did run Monte Carlo (MC) simulations to confirm the behavior predicted by the analytical model. The MC simulations are run assuming that: *i)* charge trapping and de-trapping are stochastic events governed by characteristic time constants, which are uniformly distributed on a log scale; *ii)* the number of traps is assumed to be Poisson distributed, and the average number of traps in a device is proportional to the device area; and *iii)* the $V_T$ fluctuation induced by a single trap is a random variable, exponentially distributed, being the average amplitude inversely proportional to the device area. These assumptions are in line with relevant experimental data for RTN and BTI [1, 3], as well as with TCAD analysis [9, 10]. Please note that these assumptions are done only to allow running the MC simulations. The analytical model derived in the previous section is valid for any statistical distribution of number of traps and $V_T$ fluctuation (no particular distribution is assumed in model derivation).

**Table 1**
**Values of Parameters used for the Monte Carlo Simulations**

| W x L (µm x µm) | <n> | < $\delta V_{Ti}$ > (mV) |
|---|---|---|
| 0.06 x 0.03 | 1.8 | 2.944 |
| 0.12 x 0.03 | 3.6 | 1.472 |
| 0.24 x 0.03 | 7.4 | 0.736 |
| 0.48 x 0.03 | 14.4 | 0.368 |
| 0.96 x 0.03 | 28.8 | 0.184 |

Besides illustrating the applicability of the model here developed, the results from Monte Carlo simulation are compared to the analytical model, validating it. Values for the parameters are taken from the literature. The average $V_T$ fluctuation due to a trap <$\delta V_{Ti}$> is assumed to scale inversely with transistor gate area, <$\delta V_{Ti}$>=$B_\eta$ /(W.L) [1]. For details please see equation (6) in [1] and related discussion. In the 28nm technology studied in [1], for the nMOSFET $B_\eta$ = 0.0053 mV/$\mu$m$^2$ [1]. For the calculation of average number of active traps per device <n>, a defect density of $10^{11}$/cm$^2$ is assumed [1, 8]. Table I shows the device sizes used in the Monte Carlo simulations and evaluation of analytical model equations, as well as the respective average number of active traps per device <n> and the average $V_T$ fluctuation due to a trap <$\delta V_{Ti}$>.

For each device size, 1000 Monte Carlo simulations are run. For each simulation, the size of the time series is $3x10^5$ points. At each simulation time step, trap switching probability (capture or emission of a charge carrier by a trap) is evaluated according to the trap capture or emission time constant. As an example, figure 2 shows the first $10^4$ points of the Monte Carlo runs of two transistors with *W*x*L* of 0.12µm x 0.03µm. The device in Fig. 2(a) shows a much larger $V_T$ fluctuation than the one in Fig. 2(b), illustrating the importance of properly modeling not only the expected $V_T$ jitter in a transistor, but also its variability among devices that by design should be equal. For the transistor in Fig. 2(a), the threshold voltage jitter $V_{Tjitter}$ was $\sigma_{\Delta VT} = 5.03mV$ ($\sigma^2_{\Delta V_T} = 25.4mV^2$). For the transistor in Fig. 2(b), the threshold voltage jitter $V_{Tjitter}$ was $\sigma_{\Delta VT} = 0.013mV$ ($\sigma^2_{\Delta V_T} = 0.00016mV^2$).

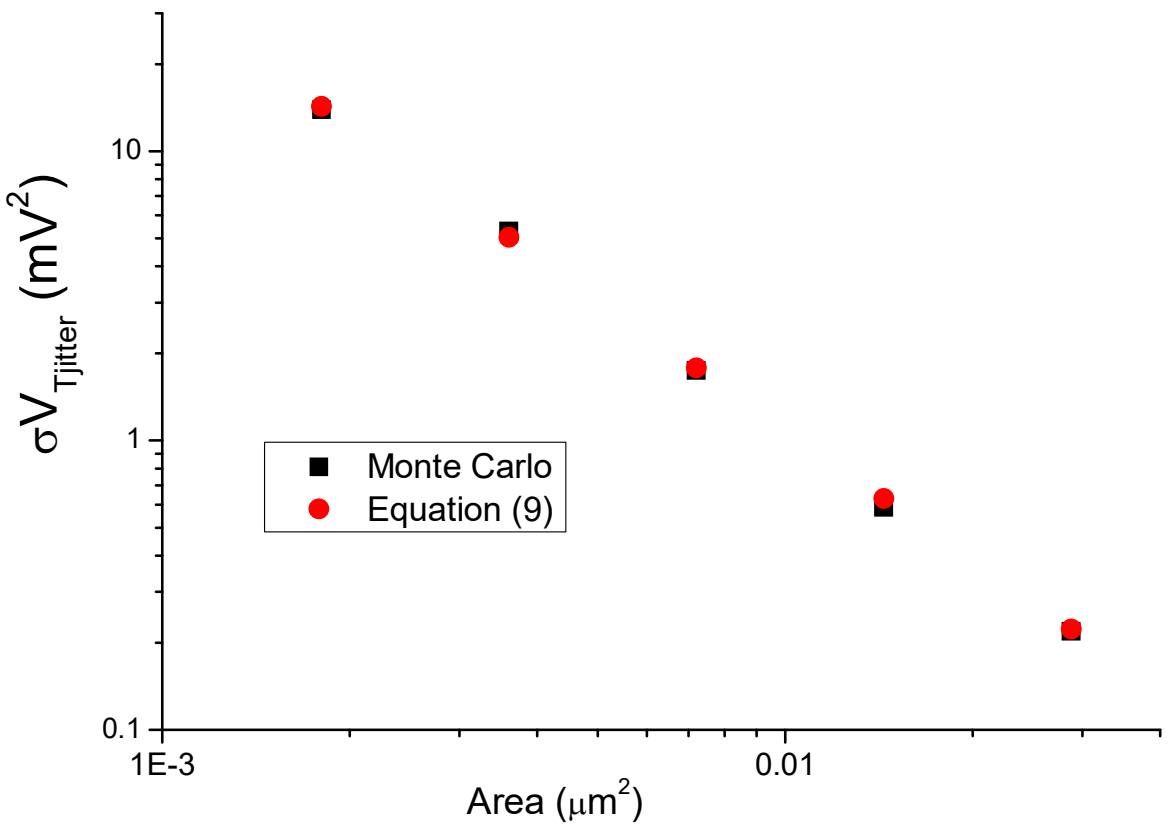

Fig. 5. $\sigma_{V_{Tjitter}}$ as a function of device area. Black squares show the value obtained from the Monte Carlo simulations. Red dots show $\sigma_{V_{Tjitter}}$ as predicted by (9).

The values obtained in the Monte Carlo simulations are compared to the analytical model. Please note that the moments of an exponentially distributed random variable $X$ are given by $<X^n> = n!/\lambda^n$, with $<X>=1/\lambda$ being the parameter of the exponential distribution [6]. This allows evaluation of equations (7) and (9) and comparison to the results from Monte Carlo simulations.

The results are shown in Fig. 3. Each point in the graph is the value for a Monte Carlo simulation run, i.e., the value of $\sigma^2_{\Delta V_T}$ of a single device. As expected, due the random number of traps, random trap amplitude and random trap activity, the $V_{Tjitter}{}^2$ is different for each device. Black diamonds show the average jitter for each device size, which is in very good agreement to value predicted by equation (7). The area scaling of $V^2_{Tjitt}$ and $\sigma^2_{V_{Tjitter}}$ is as predicted by equations (10) and (11), respectively. As expected, due to the area scaling of $\sigma^2_{V_{Tjitter}}$, in Fig. 3 the maximum vertical value for each area increases rapidly as area becomes smaller. This behavior has been experimentally observed, but analytical modeling was lacking. The RTN distributions clearly show the non-Gaussian distribution, with variation increasing as the device area scales down, as observed e.g. in [2, 4]. Furthermore, it is seen to be a heavy tailed distribution, as observed e.g. in [2, 5].

Figure 4 shows the histograms of the Monte Carlo simulations for the two different device sizes. Each histogram shows the $V_{Tjitter}{}^2$ for the 1000 devices of same size.

It is also important to address the question of how the $V_T$ jitter varies among devices that by design should be identical

(have the same W x L). This question is addressed by looking at the variance of $V_{Tjitter}^2$, which is $\sigma^2_{V_{Tjitter}}$, given by equation (9).

Figure 5 shows how the $V_T$ jitter varies among transistors that by design should be identical. This is addressed the variance of $V_{Tjitter}^2$, which is $\sigma^2_{V_{Tjitter}}$, given by equation (9). It is seen that variability increases with decreasing device size, as predicted by the model. Black squares show the value obtained from the Monte Carlo simulations. Red dots show $\sigma_{V_{Tjitter}}$ as predicted by (9). Very good agreement between analytical model and Monte Carlo simulations is seen.

This increase in variability with decreasing device size has been also experimentally observed. In [13], authors observed that by slightly increasing the transistor size, more than 50% reduction of ring oscillator frequency uncertainty can be achieved. Ring oscillator uncertainty was related to transistor delay uncertainty due to RTN.

The results also highlight the relevance of studying the shape of the distribution of the $V_T$ fluctuation ($\delta V_{Ti}$) induced by a single trap. For instance, for an exponentially distributed $\delta V_{Ti}$, $V^2_{Tjitter}$ is factorial of 2 (i.e., 2 times) larger than if compared to a constant $\delta V_{Ti}$, while $\sigma^2_{V_{Tjitter}}$ is factorial of 4 (i.e., 24 times) larger. The area scaling, however, does not depend on the shape of the $\delta V_{Ti}$ distribution.

## 4. Conclusion

In this work we extend the knowledge of the time-dependent random variability induced by RTN, by providing an analytical model for the threshold voltage jitter produced by RTN. We addressed not only the average (expected) jitter value, but also its variability among devices. The area scaling of RTN induced jitter and its variability is detailed and discussed, supporting designers in transistor sizing towards a more reliable design. Monte Carlo simulations are run, validating the analytical model and illustrating its applicability.

## References

[1] M. Simicic, P. Weckx, B. Parvais, P. Roussel, B. Kaczer and G. Gielen, "Understanding the Impact of Time-Dependent Random Variability on Analog ICs: From Single Transistor Measurements to Circuit Simulations," in IEEE Trans. on VLSI Systems, v. 27, pp. 601-610, 2019.

[2] T. Matsumoto, K. Kobayashi and H. Onodera, "The impact of RTN-induced temporal performance fluctuation against static performance variation," 2017 IEEE Electron Devices Technology and Manufacturing Conference (EDTM), Toyama, 2017, pp. 31-32

[3] G. I. Wirth, Jeongwook Koh, R. da Silva, R. Thewes and R. Brederlow, "Modeling of statistical low-frequency noise of deep-submicrometer MOSFETs," in IEEE Tran. on Electron Dev., v. 52, pp. 1576-1588, 2005.

[4] N. Tega et al., "Impact of HK / MG stacks and future device scaling on RTN," 2011 International Reliability Physics Symposium, Monterey, CA, 2011, pp. 6A.5.1-6A.5.6.

[5] S. Dongaonkar, M. D. Giles, A. Kornfeld, B. Grossnickle and J. Yoon, "Random telegraph noise (RTN) in 14nm logic technology: High volume data extraction and analysis," 2016 IEEE Symposium on VLSI Technology, Honolulu, HI, 2016, pp. 1-2.

[6] L. G. Parrat, Probability and Experimental Errors on Science, New York:Wiley, 1961.

[7] M. J. Deen, "Low-frequency noise in semiconductor devices - state-of-the-art and future perspectives plenary paper," 2017 International Conference on Noise and Fluctuations (ICNF), Vilnius, 2017, pp. 1-4.

[8] H. Reisinger, "The Time-Dependent Defect Spectroscopy," in Bias Temperature Instability for Devices and Circuits, T. Grasser, Ed. New York: Springer, 2014, pp. 75–109. ISBN: 978-1-4614-7909-3.

[9] A. Rossetto, T. H. Both, V. Camargo, D. Vasileska and G. Wirth. "Statistical Analysis of the Impact of Charge Traps in p-Type MOSFETs via Particle-based Monte Carlo Device Simulations", to be submitted to the Journal of Computational Electronics.

[10] N Ashraf and D Vasileska. "1/f Noise: threshold voltage and ON-current fluctuations in 45 nm device technology due to charged random traps". J Comput Electron (2010) 9: 128.

[11] E. Simoen, R. Ritzenthaler, T. Schram, H. Arimura, N. Horiguchi and C. Claeys, "On the Low-Frequency Noise of High-κ Gate Stacks: What Did We Learn?," 2018 14th IEEE International Conference on Solid-State and Integrated Circuit Technology (ICSICT), Qingdao, 2018, pp. 1-4.

[12] N. Mavredakis, N. Makris, P. Habas and M. Bucher, "Charge-Based Compact Model for Bias-Dependent Variability of 1/ $f$ Noise in MOSFETs," in IEEE Transactions on Electron Devices, vol. 63, no. 11, pp. 4201-4208, Nov. 2016.

[13] Matsumoto, K. Kobayashi and H. Onodera, "Impact of random telegraph noise on CMOS logic delay uncertainty under low voltage operation," 2012 International Electron Devices Meeting, San Francisco, CA, 2012, pp. 25.6.1-25.6.4.